\documentclass[preprint2]{aastex}
\usepackage[dvips]{color}
\usepackage{graphicx}
\usepackage{subfigure}
\usepackage{mathrsfs,amssymb}
\usepackage{amsmath}
\usepackage{natbib}
\usepackage{color}
\usepackage{ulem}

\newcommand{\erf}{\ensuremath{{\rm erf}}}

\newcommand{\kms}{\ensuremath{{\rm km s}^{-1}}}
\newcommand{\kpc}{\ensuremath{{\rm kpc}}}
\newcommand{\Mpc}{\ensuremath{{\rm Mpc}}}

\newcommand{\Msun}{\ensuremath{{M_\sun}}}

\newcommand{\Zsun}{\ensuremath{{Z_\sun}}}

\shorttitle{Semi-Numerical Simulation}

\begin{document}

\title{Semi-Numerical Simulation of Reionization with Semi-Analytical
Modeling of Galaxy Formation}

\author{
Jie Zhou \altaffilmark{1},
Qi Guo  \altaffilmark{1,2},
Gaochao Liu \altaffilmark{1,3},
Bin Yue \altaffilmark{1},
Yidong Xu  \altaffilmark{1},
Xuelei Chen\altaffilmark{1,4}
}

\altaffiltext{1}{National Astronomical Observatories,
Chinese Academy of Sciences,
20A Datun Road, Chaoyang District, Beijing 100012, China}
\altaffiltext{2}{Department of Physics, Institute for Computational Cosmology, University
of Durham, South Road, Durham DH1 3LE}
\altaffiltext{3}{College of Science, China Three Gorges University, Yichang 443002, China}
\altaffiltext{4}{Center of High Energy Physics, Peking
University, Beijing 100871, China}

\newpage
\begin{abstract}
In a semi-numerical model of reionization, the evolution of ionization
fraction is simulated approximately by the ionizing photon to baryon ratio
criterion. In this paper we incorporate a semi-analytical model of galaxy
formation based on the Millennium II N-body simulation
into the semi-numerical modeling of reionization.
The semi-analytical model is used to predict the production of
ionizing photons, then we use the semi-numerical method to model
the reionization process. Such an approach allows more detailed modeling
of the reionization, and also connects observations of galaxies at low
and high redshifts to the reionization history. The galaxy
formation model we use was designed to match the low-$z$ observations,
and it also fits the high redshift luminosity function reasonably well,
but its prediction on the star formation falls below the observed value,
and we find that it also underpredicts the stellar ionizing photon
production rate, hence the reionization can not be completed at
$z \sim 6$ without taking into account some other potential sources of ionization photons. We also considered simple modifications of the model with more
top heavy initial mass functions (IMF), with which the reionization
can occur at earlier epochs. The incorporation of the
semi-analytical model may also affect the topology of the HI
regions during the EoR, and the neutral regions produced by our simulations with the semi-analytical model appeared less poriferous than the simple
halo-based models.
\end{abstract}

\keywords{Cosmology: Reionization, Galaxy: Formation}

\section{INTRODUCTION}
\label{sec:intro}

The reionization of the hydrogen gas in the universe is a topic of forefront
research in recent years. The measurement of the cosmic microwave
background (CMB) polarization by the Wilkinson Microwave Anisotropy
Probe (WMAP)indicates that the reionization
occurred at $z_{reion}=10.6\pm 1.2$, assuming an instant reionization
model \citep{2011ApJS..192...16L}. We also anticipate more precise
measurement from the Planck
satellite \citep{2011MNRAS.413.1569M,2012ApJ...756L..16A}.
On the other hand, observations of high redshift quasars show the presence of
Gunn-Peterson trough at $z \sim 6$, which marked the end of the hydrogen
reionization process \citep{2001AJ....122.2850B,2002AJ....123.1247F}.
While we still do not have detailed knowledge about the nature of
the ionizing photon sources, \citet{1997ApJ...489....7R}
found that for the observed luminosity function, high redshift quasars
failed to produce enough ionizing photons to keep the universe ionized
before $z \sim 4$, thus the stars probably contributed the majority of ionizing
photons, but the fractions of different stellar populations and the quasars
are presently unknown. However, as the capability of space and ground based
optical/infrared telescopes improved, we are learning
more and more about the galaxies at the
epoch of reionization (EoR)
\citep{2010ApJ...722..803O,2010MNRAS.403..938W,2010MNRAS.409..855B,
2011MNRAS.414.1455L,2011arXiv1112.6406Y,2011ApJ...728L..22Y,
2012ApJ...752L...5B,2012arXiv1204.3641B}.
Moreover, hundreds of Gamma-Ray Bursts (GRB) have been detected by the SWIFT
and Fermi spacecraft\citep{2012MSAIS..21...54S},
many of these are at high redshifts, with the highest ones, e.g.
GRB 090429B and GRB 090423 at $z>8$, and they provide
additional information on the star formation history during the EoR
 \citep{2011MNRAS.418..500I}.

The 21cm emission from high redshift intergalactic medium (IGM) could
potentially allow us to observe the epoch of reionization directly,
providing 3-dimensional information about the evolution and morphology
of reionization process\citep{1997ApJ...475..429M}.
However, it is difficult to detect this signal
with the presence of strong foregrounds. Attempts have been made with
the EDGES experiment \citep{2010Natur.468..796B}, and
GMRT EoR search \citep{2011MNRAS.413.1174P,
2009MNRAS.399..181P}.
Several low frequency radio telescope arrays
have been or are being built to detect this signal,
including the 21CMA \footnote{http://trend.bao.ac.cn/index.html},
the Mileura Wide Field Array (MWA)\citep{2011PhRvL.107m1304J},
the Low Frequency Array
(LOFAR) \citep{2010MNRAS.405.2492H}, and PAPER \citep{2010AJ....139.1468P}.
In the future, HERA \citep{2009astro2010S..82F}
and the Square Kilometer Array (SKA)\citep{2011A&A...527A..93S}
may provide even more powerful observational probes.

Detailed modeling of reionization is required to interpret the various
observational data.  An accurate numerical
model must include treatment of gas dynamics, chemistry, feedback
processes, and especially, radiative transfer of the ionizing photons.
Many groups have conducted radiative transfer
simulations to study the EoR \citep{2000ApJ...542..535G, 2002ApJ...572..695R, 2003MNRAS.344L...7C, 2001NewA....6..359S, 2003MNRAS.345..379M, 2006NewA...11..374M, 2007MNRAS.377.1043M, 2007ApJ...671....1T, 2008MNRAS.386.1931A, 2008MNRAS.387..295A, 2009MNRAS.393.1090F, 2009MNRAS.396.1383P}.
However, such simulations have extremely high demands on the computing
resources. High resolution is required to resolve low mass galaxies, which may
contribute a significant fraction of ionizing photons, but as
the typical size of the ionized regions at the end of EoR is expected
to be tens of comoving Mpc, a large simulation box is also required to
statistically sample the distribution of HII regions. This large dynamic
range put severe demands on the computational cost of the simulation.
Furthermore, since our knowledge about high redshift galaxies
is still very rudimentary, we need to explore a large range of
parameter space, but the high computational cost make this
difficult or impossible to do. For these reasons, it is very
worthwhile to make simpler but faster models to gain some physical insights
and explore the parameter space.

Inspired by the results of numerical simulations with
radiative transfer computation of ionizing photons,
an analytical model known as the ``bubble model'' was developed
\citep{2004ApJ...608..622Z, 2004ApJ...613....1F, 2004ApJ...613...16F}.
It uses an excursion set formulation of structure formation to study the
size distribution of ionized regions and the induced 21cm emission
power spectrum.
In this model, whether a region is ionized or not is determined
by the ratio of the number of ionizing photons
produced locally and in the surrounding regions to the number of baryons.
This model is intuitive, and allows one to
calculate the size distribution of the HII regions during the EoR
analytically. This method is then extended to the so called semi-numerical
model \citep{2007ApJ...654...12Z, 2007ApJ...669..663M, 2011MNRAS.414..727Z, 2011MNRAS.411..955M}.  First order perturbation theory is used to produce the
halo distribution function at any given redshift, then the
star formation rate (SFR) and the number of ionizing photons produced is
calculated. Finally, the same ionizing photon-to-baryon ratio
criterion is used to obtain the ionization field from the halo field.
\citet{2011MNRAS.414..727Z} compared this algorithm with a ray-tracing
radiative transfer simulation, and found that they are in good agreement, even though it only cost a tiny fraction of the computing
time of the full radiative transfer simulation.
However, these semi-numerical models do not
consider the galaxy formation process in detail.

The galaxy formation process is complicated. Besides the nonlinear
evolution of dark matter density fluctuations, it also involves the heating
and cooling of gas, ionization and recombination, formation of molecules
and chemical enrichment, formation of stars and feedback process.
Limited by the dynamic range of simulations and our knowledge
about the complex physics of these processes, all numerical simulations have to
adopt some kind of phenomenological approximation. In the so called
semi-analytical
modeling approach, galaxy formation processes, esp. star formation,
is simulated by following a set of prescriptions based on the
property of the dark matter halos, and after specifying some model
parameters, it can be used to predict the observational properties
of galaxies, such as luminosity function, stellar age and metal distribution,
etc. at different redshifts. The model parameters are adjusted to fit the
observations. Comparing with hydrodynamic simulation, the major
advantage of semi-analytical model
is that it has much lower computational cost,
so a large range of parameter space can be explored without
repeating the whole simulation. Semi-analytical modeling
has become a powerful tool of
cosmological investigation, and over the years the models have been
developed to incorporate more physical details, and to provide
predictions of more observations\citep{1991ApJ...379...52W, 1993MNRAS.264..201K, 1994MNRAS.271..781C, 1999MNRAS.303..188K, 1999MNRAS.310.1087S, 2000MNRAS.319..168C, 2001MNRAS.328..726S, 2003MNRAS.343...75H, 2006MNRAS.365...11C,2006MNRAS.370..645B, 2007MNRAS.375....2D,2009MNRAS.396...39G, 2010MNRAS.406.2249W}.

Recent semi-analytical models can fit
the observational data such as the luminosity and mass distributions of
galaxies and quasars, the history of star formation and quasar evolution,
and also the correlation of a number of observable properties of galaxies
and quasars \citep{2006MNRAS.370..645B,2007MNRAS.375....2D}. Semi-analytical
modeling has also been used to study high redshift universe,
for example, \citet{2011MNRAS.410..775R} used the
Durham GALFORM model to investigate the reionization history. They
compared the ionizing photon number with the hydrogen atom number in
different models of initial stellar mass functions, and found that
the stellar components were enough to ionize the universe at $z \sim 10$.

In the present paper, we introduce the semi-analytical modeling
of galaxy formation into
the semi-numerical model of reionization. The semi-analytical model
provides more detailed
information about galaxy formation and the production of ionizing photons.
More importantly, this allows us to compare the EoR models and probes with the
observations at lower redshift, so that eventually a self-consistent model
of galaxy formation and reionization could be developed.
We use the semi-analytical model developed by \citet{2011MNRAS.413..101G},
which was based on the Millennium and Millennium II N-body
simulations \citep{2009MNRAS.398.1150B}. By modeling a number of
physical processes in a plausible way, and after tuning the parameters
this model successfully reproduced many observables, such as the
luminosity function, star formation rate etc. We then use the
density field from the same Millennium II simulation, with the baryon
density tracing the the dark matter density, and the semi-numerical
model of \citet{2011MNRAS.414..727Z} to calculate the evolution of ionization fraction.

This paper is organized as follows. In section 2 we briefly review the
semi-numerical algorithm, in section 3 we introduce the semi-analytic
model and describe our improvements. In section 4 we show our result of
the reionization history and morphology. In section 5 we discuss our
parameter space and summarize our results.

\section{Semi Numerical simulation}
\label{sec:sns}

The semi-numerical model of reionization\citep{2011MNRAS.414..727Z,
2011MNRAS.411..955M} is an extension of the
analytical bubble model \citep{2004ApJ...613....1F}. It is assumed
that before the completion of reionization process of the IGM,
a number of HII regions will appear in the IGM, and they are preferentially
located in regions of higher densities, because in such regions
structures formed earlier. The mass of each HII region is
proportional to the number of ionizing
photons produced within the region, the criterion of ionization of the
region is
$f_{coll}>\zeta^{-1}$,
where $\zeta$ is the efficiency parameter, $f_{coll}$ is the fraction of mass in collapse object. It could be written as
$\zeta=f_{esc} f_* N_{\gamma}/(1+N_{rec})$, where $f_{esc}$ is the fraction
of ionizing photon which escaped the halo into the IGM,
$f_*$ is the fraction of baryons in stars in the halo, $N_{\gamma}$ is
the number of photons emitted per baryon in the stars, and $N_{rec}$
is the average number of recombinations. We can rewrite the reionization criteria as
\begin{equation} \label{eq:detm}
\delta_{m}> \delta_{x}(m,z)
\end{equation}
where
\begin{equation}
\delta_{x}(m,z)=\delta_{c}-\sqrt{2}K(\zeta)[\sigma^2_{min}-\sigma^2(m)]^{-1/2}
\end{equation}
 and
$K(\zeta)=\erf^{-1}(1-\zeta)$, $\delta_{m}$ is density fluctuation and $\delta_{x}$ is the reionization criteria. A point in a
region of mass $m$ is marked as ionized if and only if the scale $m$ is
the largest scale for which the condition Eq.~(\ref{eq:detm}) is satisfied.
The mass function of the HII regions can then
be obtained. For normal stars, it was estimated that a plausible set
of parameters are $f_*\sim 0.2$, $N_{\gamma}\sim 3200$, $f_{esc}\sim 0.05$,
$N_{rec}\sim 2$, so $\zeta \sim 10$. However, there are large
uncertainties in all of these parameter values.
For example, most low redshift
observations indicate  $f_{esc} < 5\% $, but the escape fraction at
high redshifts might be much larger. Observations of
galaxies at $z = 1-3$ indicate a broad range
of escape fraction values from a few percent to tens
of percent\citep{2006ApJ...651..688S, 2007ApJ...668...62S,
2011A&A...532A..33G, 2009ApJ...692.1287I}. $\zeta$ is also very
uncertain, and at present vastly different choices of $\zeta$ value is
possible \citep{2004ApJ...613....1F}.

\citet{2011MNRAS.414..727Z} and \citet{2011MNRAS.411..955M} generalized the
analytical bubble model to what they called semi-numerical simulations, and the latter are publicly avaiable known as the {\tt 21cmFAST}.
The procedure for such calculation is:
\begin{flushleft}
\begin{enumerate}
\item Creating the linear density and velocity fields;
\item find halos from the density field;
\item reallocate halo position by first order perturbation theory;
\item generate ionizing field by the equation: $\zeta*m_{gal}>m_{H}$
\end{enumerate}
\end{flushleft}

In step 2 and 4, the formation criteria are checked from large scales down to
each single cell of the simulation box to flag a halo
or an HII region. Once the criterion is satisfied, a halo is generated (in
step 2) or an ionized region is marked.  For the ionization,
one could either flag all pixels inside the region as ionized, or
only flag the center pixel as ionized. Obviously, the latter is much faster, but the results turned out to be
similar with no significant difference.
\citet{2011MNRAS.414..727Z} compared their
result with a radiative transfer simulation with the same initial conditions,
and found the result of the semi-numerical simulation was a good
approximation of the radiative transfer simulation.
Thus, the semi-numerical algorithm captures the bubble topology and
reionization history quite well with moderate amount of computation.

In this paper, we further improve the semi-numerical model by implementing
a more detailed model of ionizing photon production based on the
semi-analytic model of galaxy formation. Both the semi-numerical model of
reionization and the semi-analytical model of galaxy formation are
efficient in computation, so our model still allows relatively quick exploration
of the parameter space. Moreover, this approach also allows us to
investigate how the physical processes affect reionization history, and to
constrain the reionization model parameters with galaxy observations.

To include the physical process in the galaxies, we rewrite the
ionizing criteria as
\begin{equation} \label{eq:crt}
\frac{N_{\gamma} f_{esc}}{1+N_{rec}} > {N_{IGM}},
\end{equation}
where $N_{IGM}$ is the hydrogen number density in the IGM.
The Millennium II simulation is a pure dark matter simulation,
we assume that the baryon density traces the dark matter density,
$\rho_{b}=\rho_{d}*\Omega_{b}/ \Omega_{m}$. For the reionization simulation,
we smooth the density field on to $256^3$ grid.
We calculate $N_{\gamma}$
for each galaxy by relating its UV luminosity with the
star formation rate, and by integrating this luminosity
through its formation history we could get the
total number of ionizing photons. Our algorithm is:
\begin{flushleft}
\begin{enumerate}
\item convert dark matter density field in Millennium II
simulation to IGM density field;
\item located galaxies from semi-analytical model into simulation box and
calculate ionizing photon number;
\item generate the ionizing field by Eq.(\ref{eq:crt}).
\end{enumerate}
\end{flushleft}

In step 2, we assume that for ordinary Pop I stars the total number of ionizing
photons is similar to the value given in
\citet{2004ApJ...613....1F}. However, we also tested how the stellar population
affects the ionizing history. In our model C (see next section)
we treat the luminosity as a function
of metallicity. In step 3, for each pixel we use the same method
as \citet{2011MNRAS.414..727Z}, checking the ionizing criteria from
a large scale comparable with the simulation box and step downwards.
Once the criteria is satisfied we flag all pixels inside the region
as ionized, or if not, we move into a smaller radius to repeat the
above process, till the region is ionized or we have reached a
single pixel.

The effect of recombination at high redshift is quite complicated, since
the purpose of this paper is to develop a relatively fast
method to study the ionizing history and the morphology of HII regions, here we
shall make a simple assumption of a constant number of recombinations
here (see \citealt{2012ApJ...747..127Y} for an example of
sophisticated modeling of the evolution of recombination rate).
Here we adopt $N_{rec}=2$.

\section{Semi-Analytical Model}
\label{sec:sam}

In the hierarchical structure formation scenario, the dark matter halos
grow by accretion and merger, and galaxies form within dark matter halos
by the radiative cooling of the baryonic gas. In semi-analytical models,
one follows the evolution of each dark matter halo, and apply a set of
rules to describe the gas cooling, star formation and feedback for each
halo without actual simulations. The properties of the galaxies within each
halo, such as the stellar mass, the age of stellar population, distribution
within halos, and the amount of cold and hot gas and metal abundance can be
followed.  The dark matter halo merger tree is generated either by Monte-Carlo
simulations\citep{1991ApJ...379...52W,1993MNRAS.264..201K,1994MNRAS.271..781C},
or by using an N-body simulation.
With improvement in computing power,  in most
recent works the latter approach is adopted \citep{2006MNRAS.370..645B,
2006MNRAS.365...11C,2007MNRAS.375....2D,
2009MNRAS.396...39G, 2010MNRAS.406.2249W}.

For galaxy formation modeling during the epoch of reionization, high
resolution simulation is required. Indeed, it is believed
that the first stars form in halos of $10^6 \Msun$, and
the first galaxies have masses about $10^8 \Msun$
(c.f. \citealt{2001PhR...349..125B}). At the same time, the model
should also contain enough volume such that a large number of
neutral or ionized regions can be included in the simulation at the EoR.
In this work, we use a model based on the Millennium II simulation (MS II)
\citep{2009MNRAS.398.1150B}. This is an
extension of the earlier Millennium simulation \citep{2005Natur.435..629S},
and is among the largest cosmological N-body simulations with
sufficient mass resolution that is currently available.
It assumed a $\Lambda CDM$ cosmology with parameters based on
combined analysis of the 2dFGRS \citep{2001MNRAS.328.1039C} and the
first year WMAP data \citep{2003ApJS..148..175S}. The parameters
are: $\Omega_{m}=0.25$, $\Omega_{b}=0.045$, $\Omega_{\Lambda}=0.75$,
$n=1$, $\sigma_8=0.9$ and $H_0=73 \kms \Mpc^{-1}$. These parameters
differs slightly from the more recent best fit values\citep{2011ApJS..192...16L}
but the relatively small off-set
are not significant for most of the issues discussed in this paper, as there
are much larger uncertainties in star formation and reionization parameters.
There are $2160^3$ particles in MS II, the box size is $100 \Mpc/h$, the softening length $1\kpc/h$, and
the particle mass is $6.9*10^6M_{\odot}/h$.

\citet{2011MNRAS.413..101G} developed a semi-analytical model
based on the MS II. This model is an extension and improvement
of earlier models based on
the Millennium simulation \citep{2005Natur.435..629S,2006MNRAS.365...11C,2007MNRAS.375....2D}. In this model, central galaxies, satellite galaxies in subhalos
and orphan galaxies which had lost their subhalos are distinguished.
The baryonic content of the galaxies is split into
five components, including a stellar bulge, a stellar disk, a gas disk, a hot
gas halo, and an ejecta reservoir. In addition, intra-cluster light is
also included. The model keep track of various
processes involved in galaxy formation, including gas heating
and cooling, evolution of the stellar and gas disks, star formation, supernova
feedback, gas striping in groups and clusters, merger and tidal disruption,
bulge formation, black hole growth and AGN feedback, metal enrichment,
and photo-heating of the pregalactic gas by the UV background after
reionization. A Chabrier IMF
\citep{2003PASP..115..763C} is adopted, this IMF
has fewer low-mass stars than the Salpeter IMF. The model use the
stellar population synthesis model of \citet{2003MNRAS.344.1000B},
and the dust extinction model developed in \citet{2009MNRAS.396...39G}.
It predicts observable properties such as the stellar mass function,
mass-size relation, distribution of galaxies within cluster and
group halos, morphological type, black hole mass -- bulge mass relation,
low redshift galaxy luminosity function for different observing bands,
stellar mass-halo mass relation, cold gas metallicity, galaxy color
distribution, Tully-Fisher relation, satellite luminosity function,
autocorrelation function for different masses, etc. Some of these predictions
are compared with observations to fix the model parameters.
Compared with earlier semi-analytical model \citep{2007MNRAS.375....2D},
which overpredicts the abundance of galaxies with mass near
or below $10^9 \Msun$ when applied to the MS II, this model
improved the treatment of a number of physical processes, including
the treatment of supernova feedback, reincorporation of ejected gas,
sizes of galaxies, treatment of satellite galaxies which are outside
the $R_{200}$ but belongs to the same group, and
environmental effects. The model is adjusted to best
fit the observational data on galaxies at $z \sim 0$, as
there are much more high quality
observational data available at low redshifts, and they are subject
to less selection effect.  In this paper we adopt this
model as our {\it model A}
and apply it to the semi-numerical simulation of reionization.

While the predictions of this semi-analytical model are in good
agreement with many observations at $z \lesssim 1$, and the
abundance of luminous galaxies are also consistent with observations
at $z \lesssim 3$,
the predicted high redshift star formation rates are systematically
lower than the observed values (see Fig.22
of \citealt{2011MNRAS.413..101G} and discussions there).
This is not simply a problem with this particular model, for it is well
known that if the observed star formation rate is integrated over redshifts,
the luminosity function of galaxies would be
overpredicted (c.f. \citealt{2008MNRAS.385..687W}).
In fact, \citet{2011MNRAS.413..101G} argued that even the present model
might have overpredicted the abundance of low mass galaxies at high redshifts.
However, as we shall see in the next section, with this
model the ionizing photon
production rate might be too low to allow the universe to be
 reionized at $z > 6$.

There have been other semi-analytical studies which tries to accommodate more
of the high redshift SFR observations. For example, in most semi-analytical model it is assumed that major mergers can trigger starbursts, so that most gas in the merging galaxies
collapse and form stars in a short time. \citet{2011MNRAS.410..775R} argued that in such a scenario the stars are much more heavy than stars formed continuously. So they assumed a top heavy initial mass function for the stars formed in major mergers, and the number of ionizing photons would be increased dramatically.  Major merger are more frequent
at high redshifts, so this model can be expected to increase
the production of ionizing photons by a factor of 5 to 10 during the EoR.

To achieve a higher photon production rate during the EoR, we may consider
a more top heavy IMF as \citet{2011MNRAS.410..775R} did. Strictly speaking,
once a semi-analytical model is set, its model
assumptions or parameters such as the IMF should not be altered,
because the feedback
would change subsequent star formation, and if one runs the full
semi-analytical model with the new assumptions and compare the results
with observations, a different set of parameters would be obtained. However,
as we noted above, at present there is a conflict between the low redshift
abundance and high redshift star formation rate, which can not be
easily solved. So to illustrate the effect of change,
here we will just make a simple modifications of the
IMF, and examine its impact without considering the feedback.
In our {\it model B}, we make an assumption similar to
\citet{2011MNRAS.410..775R}, that major merger will trigger star burst,
the IMF in star burst galaxies are very top heavy with an
index of 0, mass range of $(0.15-120 M_{\odot})$. By doing this, more than half of the
newly formed stars are massive ones, and
the number of ionizing photon is about 10 times
than {\it model A}.

As the model B may be too extreme, we also consider a
{\it model C}. As noted above,
during the epoch of reionization, many of the stars are
formed in metal free or very low metallicity regions,
which may have a top heavy IMF. Furthermore, even for the same mass,
the metal poor stars produce more ionizing photons. Here, we do not assume
the flat IMF as in model B, but instead consider a more moderate model
as suggested in \citet{2003A&A...397..527S}, who examined the spectral
properties of the ionizing continua, the Lyman break and
recombination lines in star bursts and constant star formation phase,
for metallicity from 0 up to solar metallicity. We adopt a Salpeter IMF for all galaxies, attributing very massive stars in the range of $(1-500 M_{\odot})$ with index of 2.3, and mass range of $(1-100 M_{\odot})$ with index of 2.55 for the galaxies with higher metallicity. Applying these results, the procedure of our simulation with model C is
\begin{flushleft}
\begin{enumerate}
\item The ionizing luminosity for a galaxy at a given redshift
is determined by its metallicity, which is determined by the
metallicity of the cold gas at last redshift.
\item We interpolate the property of the
stellar population from \citet{2003A&A...397..527S} and
\citet{2003MNRAS.344.1000B}, to obtain the luminosity;
\item Integrate the luminosity in the past to get the number of ionizing photons;
\item Generate ionizing field with the number of
ionizing photons from step 3 and IGM density.
\end{enumerate}
\end{flushleft}

\begin{figure}[t]
\epsscale{1}
\plotone{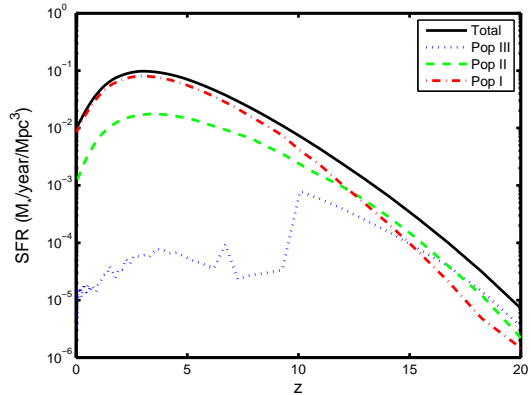}
\caption{The SFR of different metallicities in our model C. Here the criteria of Pop II and Pop III stars are $\frac{1}{50}\Zsun$ and $10^{-4}\Zsun$ respectively.
\label{fig:SFRfrac}}
\end{figure}

We may estimate the mean metallicity of the newly formed stars from the
the metallicity in the cold gas of the halo. However, in the original
semi-analytical model, newly formed halos (i.e. without progenitors),
regardless of redshift, will always be metal free, as the metals
are not transferred out to the IGM in that model.
In reality, the IGM will be polluted by galaxy winds, which
blow the enriched gas out of the halo, and Lyman alpha forest observations
show that the IGM has already been contaminated at $z>3$ \citep{2003ApJ...596..768S}.
Thus, even the newly formed halos should contain some amount of metals.
As a remedy of this, we set the
metallicity in the cold gas of the newly formed halos as the
average value of the whole simulation box, which is calculated by the
metal production in the original semi-analytical model.
In Fig.~\ref{fig:SFRfrac},
we plot the star formation rate of different metallicities according to this
model, the Pop I, II, and III stars are represented by the blue dot, green
dash and red dot-dash curves, respectively, and the black solid curve represents
the total SFR. While initially (at very high redshift )
the Pop III had the largest SFR,
this was soon surpassed by Pop II ($z \sim 16$),
which was in turn surpassed by Pop I at a fairly high redshift ($z \sim 12$).
The formation of Pop III stars declined at $z\sim 10$, and fluctuates a little
at low redshift but stays at low value. In this model we have assumed
the metals are uniformly distributed, the non-uniform distribution of metals
may allow a higher production rate of metal free or metal poor stars.

\begin{figure}[t]
\epsscale{1}
\plotone{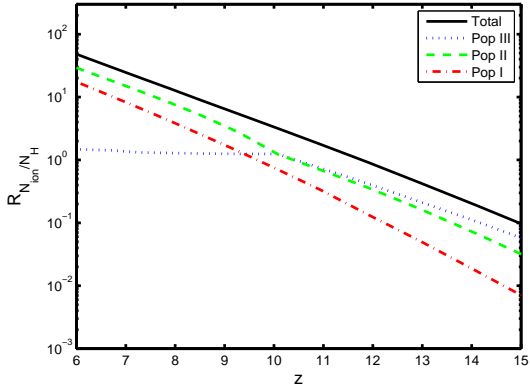}
\caption{The number of ionizing photons of different metallicities in our model C. Here the criteria of Pop II and Pop III stars are $\frac{1}{50}\Zsun$ and $10^{-4}\Zsun$ respectively.
\label{fig:IonHist}}
\end{figure}

In Fig.~\ref{fig:IonHist},
we plot the evolution of the ionizing photon cumulative production
rate due to stars of different metallicities in our model C.
We see that during most of EoR,
the contribution from the Pop II stars dominates. The Pop III stars could make
up major contribution only at fairly high redshifts, but levels off at
$ z \sim 10$ in this model. The Pop I stars have a higher SFR, but
do not make up the major contribution to the ionizing photons.

\section{Results}

\begin{figure}[t]
\epsscale{1}
\plotone{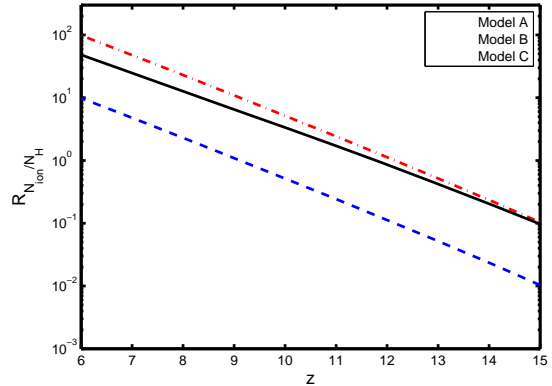}
\caption{The ratio between number of ionizing photons from
stars and number of hydrogen atoms for the three models
we considered. \label{fig:ionnum}}
\end{figure}

Let us first take a look at the global production rate of
ionizing photons in our model. In Fig.\ref{fig:ionnum},
we plot the ratio between the number of ionizing photons from
stars and the number of hydrogen atoms per comoving volume, here
the escape fraction and recombination number are not included.
Thus, if we assume $f_{esc} \sim 0.15$ and $N_{rec}\sim 2$, the
ionizing photo-to-baryon ratio has to be greater than 20 for the
Universe to be reionized.
The blue dash line shows the result for assuming
that each hydrogen atom in the stars produces on average
3200 photons during the life time of the star, which corresponds to
the normal stellar population given by our model A.
The red dot dash line shows the result for our model B,
which produce much more photons as we assumed a very top heavy IMF for
starburst galaxies. The black solid line corresponds to our model C,
which assumed a mixture of Pop I, Pop II and Pop III stars.
From the figure, we can see that for model A,
it is very difficult to ionize the universe by $z\sim 6$, even if
large escape fractions and low recombination rates are assumed. Model B
could ionize the universe at relatively early time, perhaps $z \sim 8$.
In model C, reionization occurred at $z \sim 7$, which is still much
later than the $z_{re}=9$ given by the WMAP
data \citep{2011ApJS..192...16L}.

\begin{figure}[t]
\epsscale{0.9}
\plotone{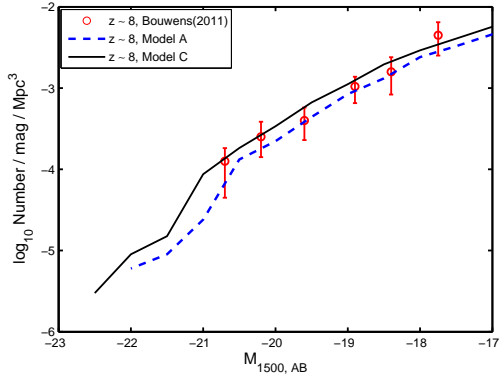}
\caption{The luminosity function at $z=7.88$. The data points are
from \cite{2011ApJ...737...90B}. The prediction of our model A and
model C are shown as  blue dash line and black solid line respectively.
\label{fig:uvlum8}}
\end{figure}

\begin{figure}[t]
\epsscale{0.9}
\plotone{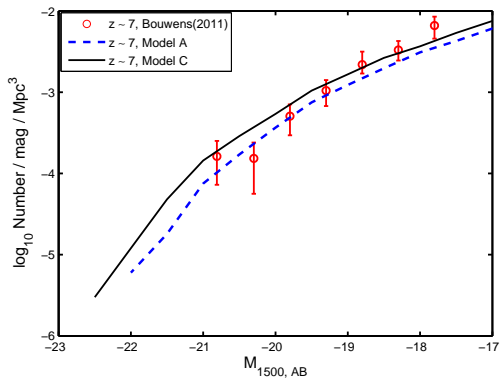}
\caption{The same as Fig.~\ref{fig:uvlum8}, but for $z=7.27$.
\label{fig:uvlum7}}
\end{figure}

We may also compare the galaxy UV luminosity function
with observations at high redshifts. Figs.~\ref{fig:uvlum8}-\ref{fig:uvlum7}
show the 1500\AA~  luminosity functions predicted by our model A and model C
at $z=7.88$ and $z=7.27$ respectively,
and for comparison we also plot the measured luminosity function
at 1600\AA~  from \cite{2011ApJ...737...90B}. We see that our model A
is in excellent agreement with the luminosity function, even though its
prediction on SFR falls below many observations.
Our model C overpredicts the luminosity function slightly, but given the large
uncertainties in the data at present, the deviation is still acceptable.


\begin{figure}[t]
\epsscale{1.1}
\plotone{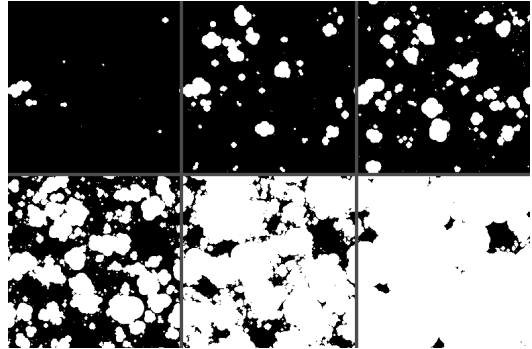}
\caption{Ionizing map of our model C, with parameters $f_{esc}=0.15$
and $N_{rec}=2$. From upper left and bottom right, the red shift are 14.9,
11.9, 10.9, 9.3, 7.88 and 7.27. Black and white region corresponding
HI and HII region, the pixels could only be total neutral or total ionized.
\label{fig:ionfield}}
\end{figure}

With the semi-analytical model of ionizing photon production in place,
we can now use the procedure described in Sec.\ref{sec:sns} to simulate
the reionization process. Figure \ref{fig:ionfield} shows the reionization
process of our simulation box for our model C.
The black colored regions are neutral, while the white ones are ionized.
The redshifts are from the upper left to bottom right 14.9, 11.9, 10.9,
9.3, 7.88 and 7.27 respectively. We can see that the universe is mostly neutral
before redshift 12, but with some bubbles of HII region.
Around redshift 9, the bubbles began to overlap, and by
$z=7$, most of the universe is ionized.

\begin{figure}[t]
\epsscale{1.1}
\plotone{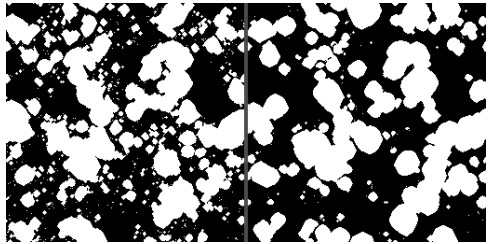}
\caption{Comparison of ionization map produced by the semi-numerical simulation
without (left) and with (right) semi-analytical models.
\label{fig:comparemap}
}
\end{figure}

How does this semi-numerical model compare with a model without the
semi-analytical modeling? This depends somewhat on the particular model
we use as well as the redshift. However, in some cases the difference
can be quite obvious. In Fig.\ref{fig:comparemap} we plot the
ionization at redshift $z=6.7$ for our model A. We have here chosen to
show model A for comparison because model A uses the original semi-analytical
model. As we can see from the figure,
while the large scale distribution of the HI and HII regions are similar,
we see that in the model without semi-analytical modeling, there are more small
HII regions which are absent in the model with it.
In the model without semi-analytical galaxy formation, the
number of photons are predicted from the dark matter halos, as a result,
some halos which do not contain stars were also assumed to produce ionizing
photons. This would not affect large HII regions, but in the neutral regions,
small HII regions are produced around these small halos. This can significantly
affect the topology of the ionization map, giving it a more poriferous
appearance. In the model with semi-analytical modeling, on the other hand,
only galaxies contribute to ionizing photons, while the starless minihalos
do not contribute, so the HI and HII region are more well separated with
a monolithic appearance.

\begin{figure}[t]
\epsscale{1}
\plotone{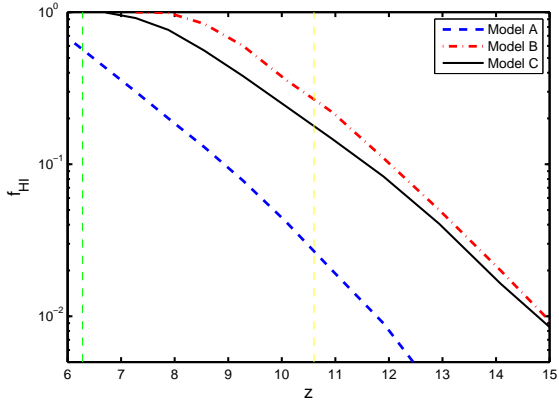}
\caption{Evolution of the ionization fraction for the
different models, here we accept $f_{esc}= 0.15$ and $N_{rec} = 2$. The vertical green and yellow lines mark the reionization redshifts constrained by Gunn-Peterson trough of quasars and WMAP observations.
\label{fig:ionhist}}
\end{figure}

In Fig.\ref{fig:ionhist} we plot the evolution of the global
ionization fraction for the different models, where we
assumed $ f_{esc}= 0.15$ and $N_{rec} = 2$.
These results are consistent with the ratio between ionizing number and
hydrogen number in figure \ref{fig:ionhist}. For model A, the reionization
happened very late, even at $z=6$, the universe is still not ionized in
this model. In model B, the ionization occurred at $z\sim 8$. In model C,
the universe is fully
ionized at $z \sim 7$, and it reach $f_{HII}=0.5$ at $z \sim 9$.

\section{Summary and Discussion}

In this paper we incorporated a Millennium II based semi-analytical
model of galaxy formation \citep{2011MNRAS.413..101G} in the
semi-numerical simulation of reionization, which was shown to
be a good approximation to the radiative transfer computation, but with
much less computational cost. Star formation rate is given by the
semi-analytical model, it is then used to compute the production rate of
ionizing photons in each galaxy. As this semi-analytical model is exclusively
designed to best fit the observations at $z\sim 0$, there is some
discrepancies between its predictions and high redshift observations.
This model, denoted our model A,
predicts an SFR lower than the observed values at high
redshifts, but it does reproduce the observed UV luminosity function
at $z \sim 7-8$, thus clearly revealing the conflict between the
observed luminosity function and SFR at high redshifts. However, the ionizing
photon production rate for this model is too low, such that the
Universe can not be reionized at $z>6$
for an escape fraction of $f_{esc}=0.15 $ and the mean
number of recombination $N_{rec}=2$.

To remedy this problem, we also considered two simple revised models.
These models assume the same star formation rate as the original model,
but adopted different IMF for star formation. In model
B, a flat IMF is assumed, this model can produce about 10
times of the ionizing photons of model A, and the universe is reionized at
$z=8$. In model C, we assumed a top heavy IMF but with similar slopes, and
we also consider the production rate of ionizing photons as a function of
metallicity of the newly formed stars. The metallicity of the newly formed stars is
determined by the metallicity of the cold gas in the halo, which was given
in the original semi-analytical model. We also test other definition of metallicity, for example, the ratio between the increase of the total mass of metals in stars and the total mass of new formed stars, i.e. the averaged metalicity in new formed stars. This definition reduces the total amount of the photons by 10\% - 20\% but won't affect our main conclusions. And we assumed a metallicity floor
in the IGM as given by the average metallicity of the simulation box. This
model predicts that Pop II stars are the major sources of ionizing photons
during the EoR, even though the Pop I SFR is higher. The predicted UV luminosity
function at  $z \sim 7-8$ of model C is only slightly higher than the observed
value. The universe is ionized at $z \sim 7$ in this model.

Armed with the semi-analytical model, we simulate reionization
history with the semi-numerical method. The addition of the
semi-analytical model could significantly affect the topology of the
neutral and ionized regions. While it is somewhat dependent on the
particular model and redshift, in some cases the difference
is quite large. Without the semi-analytical model of galaxy
formation, we may incorrectly assume ionizing photons were produced by
starless minihalos in HI regions, and this may produce a more poriferous
appearance. With the semi-analytical model, the ionizing photons come only
from galaxies, not from all minihalos, so the HII regions appeared more
monolithic and well separates from HI regions.

There are some simplifications we made in this research. For example, we have
adopted simple constant mean value for escape fraction and number of
recombinations. In reality, both of these may evolve with redshift,
or even vary with different galaxies.  We have
assumed a simple two phase model of IGM, in any given point
it is either fully neutral or fully ionized, partial ionization
is not consider, nor do we consider the effect of quasar formation.
However, within the plausible range of the parameters, these effects
are unlikely to change our result qualitatively.
As the aim of this paper is to illustrate the use of
semi-analytical model of galaxy formation in semi-numerical
modeling, we do not try to model these complications.

There are certain limitations in our present work. Perhaps the most
important one is that our model B and C are not completely self-consistent.
We have assumed that they have the same star formation rate evolution
as given by model A, but with modified IMF and ionizing photon production.
However, once the IMF of the model is changed, the star formation history
would not be the same, because the feedback effect would be different, and
that would change the subsequent accretion and star formation. If we hope
to have a similar star formation history with this modified IMF,
we must assume a lower strength of feedback.  Furthermore, the
reionization itself may also affects the galaxy formation process.
To be really self-consistent, one has to modify the semi-analytical
model itself. This will be the next step of our research.

As we discussed above, the semi-numerical simulation combined with
semi-analytical galaxy formation model provide a good approximate
of reionization history. It allows us to investigate large parameter ranges
in short time, and more importantly, connects the observations on galaxies
to the reionizations. This approach could help us better understand the
galaxy formation  process.

\section*{Acknowledgments}
We thank Youjun Lu for helpful discussions, and Qi Guo acknowledges support from the National basic research program of China (973 program under grant No. 2009CB24901), the Young Researcher
Grant of National Astronomical Observatories, CAS, the NSFC grants program (No. 11143005), and the Partner Group program of the Max Planck Society.
This work is supported by the MOST 863 Project No. 2012AA121701,
NSFC grant 11073024, CAS grant KJCX2-EW-W01, and the John Templeton Foundation.

\bibliographystyle{hapj}
\bibliography{refe}

\end{document}